\documentclass[11pt]{article}
\usepackage{amssymb}
\usepackage{graphics}
\usepackage{epsfig}
\usepackage{url}
\usepackage{hyperref}

\newcommand{\code}[1] {{\tt #1}}

\newcommand{\ket}[1]{{|#1 \rangle}}
\newcommand{\braket}[2]{{\langle #1 | #2 \rangle}}
\newcommand{\measure}[1]{{ | #1 |^{2}}}

\newcommand{\ZZ}{{Z\!\!\!Z}}

\newcommand{\R}{{\bf R}}
\newcommand{\U}{{\bf U}}

\newcommand{\INRIA}{{\href {http://www.inria.fr}{INRIA}}}
\newcommand{\INDES}{{\href {http://www-sop.inria.fr/indes}{INDES}}}
\newcommand{\PARTOUT}{{\href {http://www-sop.inria.fr/indes/PARTOUT}{PARTOUT}}}

\newcommand{\plusmod} {{\oplus}}
\newcommand{\assign} {{~:=~}}

\title {\bf Simulation of Quantum Mechanics\\
Using Reactive Programming\footnote{\small with support from ANR,
project {\PARTOUT} ANR-08-EMER-010}}

\author{
     \href {http://www-sop.inria.fr/members/Frederic.Boussinot} 
              {\Large \sc Fr\'ed\'eric Boussinot} \\
       frederic.boussinot@inria.fr \\
\\
       {\INRIA} / {\INDES} \\
       06902 Sophia-Antipolis Cedex, France
}

\date{}

\begin{document}
\maketitle
\begin{abstract}
  We implement in a reactive programming framework a simulation of
  three aspects of quantum mechanics: self-interference, state
  superposition, and entanglement. The simulation basically consists
  in a cellular automaton embedded in a synchronous environment which
  defines global discrete instants and broadcast events. The implementation
  shows how a simulation of fundamental aspects of quantum mechanics can
  be obtained from the synchronous parallel combination of a small number of
  elementary components.
\end{abstract}



\section{Introduction}

The work described here is an exercise in parallel programming inspired
by Physics (more precisely, by R. Penrose's books
\cite{Penrose-NewMind,Penrose-Road}). Actually, it presents a simulation
intended to reflect three rather intriguing aspects of Quantum Mechanics (QM):

\begin{enumerate}

\item {\it Self-interference}: this is basically Young's
  experiment where a particle interferes with itself when passing through
  two slits.

\item {\it Superposition}: the state of a particle is a superposition of
several basic states which disappears when a measure occurs.

\item {\it Entanglement}: a measure performed on one element of a pair
  of entangled particles has an instantaneous effect on the other.

\end{enumerate}

The main objective of the simulation is to build global behaviours as
{\it parallel compositions} of elementary ones. The variant of
parallelism used is the {\it synchronous} one
\cite{Halbwachs-Synchrone}. The goal is not to exactly model reality
(actually, one gets something which is is {\it certainly not} a good
model of reality as it does not compute over probabilities) but to
mimic some aspects of it, linked to QM.

We adopt Penrose's terminology and call {\U} and {\R} the two basic
procedures of QM.  The {\U} procedure corresponds to
the evolution of a system described by a state $\ket{\psi}$ which
evolves according to the rule:
\begin{equation}
\nonumber
i\hbar \frac{d}{dt} \ket{\psi}= {\cal H} \ket{\psi}
\end{equation}
This rule is deterministic and linear (any linear combination of
solutions is also a solution).
The {\R} procedure corresponds to a change of state of $\ket{\psi}$ of the form:

\begin{equation}
\ket{\psi} = \sum_{a}ˆ{} \ket{a} \braket{a} {\psi} \rightarrow \ket{a}
\end{equation}
The superposition state $\ket{\psi}$ becomes the basic state
$\ket{a}$; the probability of this change is $\measure
{\braket{a}{\psi}}$. The rule {\R} is nondeterministic and
probabilistic.


We make the assumption of a discrete world built over a cellular
automaton (CA) \cite{ToffoliMargolus-CA}, implementing the {\U}
procedure. We do not model fields, which are non-discrete objects, but
CA's which are discrete ones. Actually, in CA's, both space and time
are discrete.  The rate at which a CA is simulated is the limit
speed\footnote{it can be considered as the ``light speed'' of the system.}
of the system. CA's are embedded in a synchronous world which is
actually the one of the simulation. There are thus {\it global
  instants} during which actions are considered as simultaneous. The
basic hypothesis is that the {\U} procedure is deterministic and local
(thus, not instantaneous, as it can take several instants for a change
to propagate in space). On the opposite, the {\R} procedure is
nondeterministic, global, and instantaneous (in a sense to be made
precise later).  The main basic notions are: synchronous parallelism,
determinism of {\U} and nondeterminism of {\R}, discrete space (CA)
and discrete time (synchrony hypothesis).

In the simulation, a (virtual) particle at instant $t$ is represented by a
set of cells living at instant $t$. Each cell is in a basic state of the
particle, and the particle global state is the superposition of all the
basic states of the associated cells. Basic states are identified by
specific colors and cells are drawn according to the basic state they
hold. When superposition disappears, after execution of procedure {\R}, one
cell is chosen randomly and the created (real) particle falls in the
basic state of the chosen cell (the particle then receives the color
of the cell).  Thus, the probability for a particle to fall in a given
basic state $\ket a$ depends on the number of cells holding this state: the
more there are cells holding $\ket a$, the more the probability to
choose $\ket a$ is high.

A measure is implemented by the emission of a broadcast event that is
instantaneously received by all the cells implementing the measured
object. This is basically the {\R} procedure. In reaction to {\R}, one cell
is nondeterministically chosen and the real particle is produced from
that cell (with its basic state).

A particle is virtual when it is implemented by the CA. It turns to
real after application of the {\R} procedure. A real particle is animated
by the synchronous simulation; it is a data structure with coordinates
and speed, animated by several elementary parallel behaviours (for
example, an inertia behaviour, which at each instant sets the
coordinates according to the speed, and a bouncing behaviour which
makes the particle bounce on the simulation borders\footnote{borders
  do not have any physical significance; they are there just to contain 
  particles in a limited area.}.

The simulation basically shows the following aspects:
\begin{enumerate}
\item Self-interference is illustrated by Young's
  experiment where a particle is emitted  against a
  wall with two slits. The particle passes through both slits, which
  produces interferences. The simulation shows the production of
  interferences, and their disappearance when one slit is obstructed.

\item The simulation shows the production of a real particle when a
  detector reacts to the presence of a living cell; in this case, the {\R} procedure
  entails the reduction to a basic state.

\item Pairs of entangled particles are considered. The measure of one
element of the pair triggers the {\R} procedure for it, which produces a
real particle. The measure also triggers the {\R} procedure for the
second particle. Moreover, it forces the choice of the basic state of the real
particle produced from the second virtual particle.
\end{enumerate}

Section \ref{section:simulation} describes the simulation.  Section
\ref{section:implementation} describes the implementation and gives the
most important pieces of code.  Some related work is described in
Section \ref{section:related-work}. Finally, Section
\ref{section:conclusion} concludes the text.


\section {The Simulation \label {section:simulation}}

We consider particles whose states are superpositions of {\it
  basic states} which are integers modulo $k$ ($\ZZ /k \ZZ$); in the
sequel, $k$ is called the {\it base}, and we arbitrarily set it to
6. Thus, there are 6 distinct basic states:

\begin{equation}
\#\ket {a} = 6
\end{equation}

\begin{figure}[!htbf]
\begin{center}
\includegraphics[width=300pt]{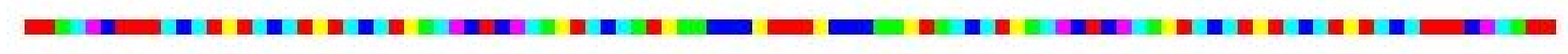}
\end{center}
\caption{Superposition State}
\label{superposition}
\end{figure}

Each basic state is assigned a distinct color. A {\it superposition}
is a (finite) sequence of basic states. Figure \ref{superposition}
shows the representation of a superposition.

\begin{figure}[!htbf]
\begin{center}
\begin{tabular}{cc}
\includegraphics[width=220pt]{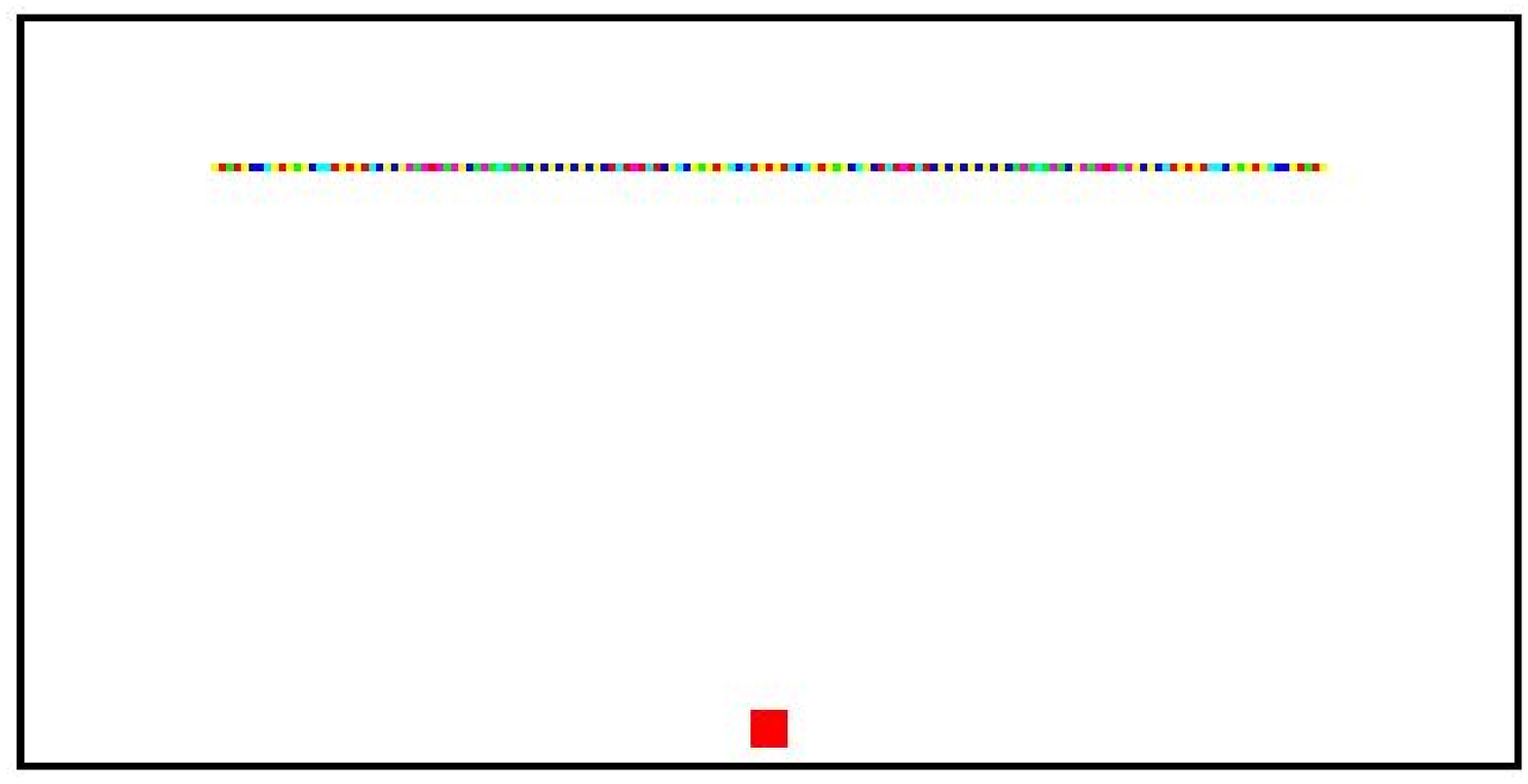}&
\includegraphics[width=220pt]{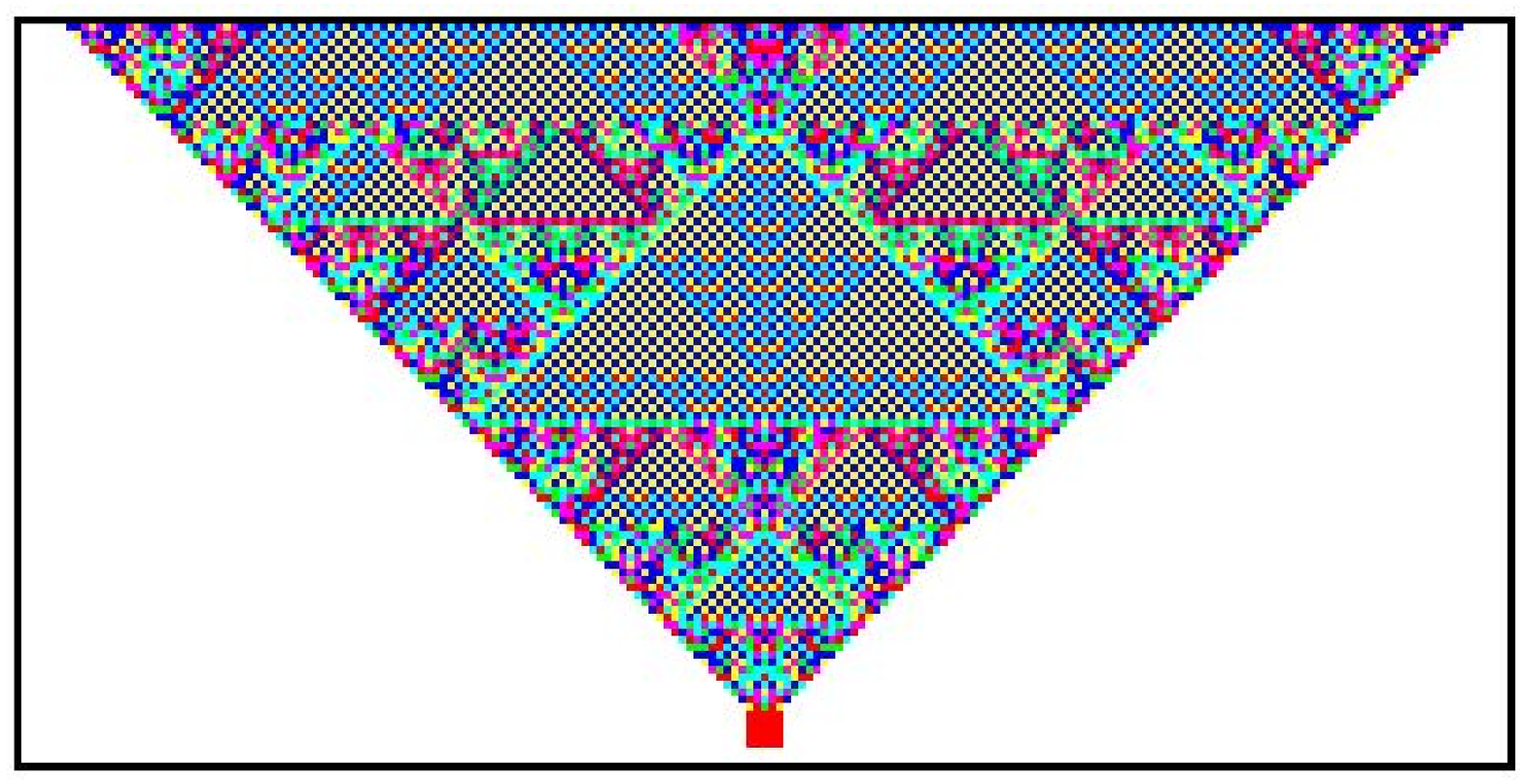} 
\end{tabular}
\end{center}
\caption{Particle Emission}
\label{one-particle}
\end{figure}

The emission of a particle by a source is represented in Figure
\ref{one-particle}.  The source is at the bottom of the figure (big red
square). The simulation is enclosed by a wall (black cells) that
forces the particle to stay inside the delimited area.  The particle is
emitted in direction of the top. Its superposition states are shown,
as time passes. The left part of the figure shows one instant, before
the particle reaches the wall. In the right part are collected all the
traces of the particle until it reaches the wall; the image obtained is
typical of a cellular automaton (this point will be discussed later).

\begin{figure}[!htbf]
\begin{center}
\includegraphics[width=300pt]{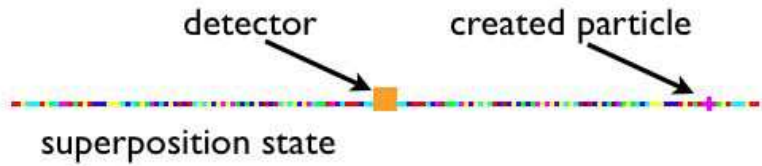}
\end{center}
\caption{Detection}
\label{detection}
\end{figure}

{\it Detectors} are special cells that react when a superposition state
reaches them. Figure \ref{detection} shows a detector (big orange
square) reacting to a superposition state reaching it. A particle is
created on the right side (with a basic state corresponding to the
pink color).

\begin{figure}[!htbf]
\begin{center}
\begin{tabular}{cc}
\includegraphics[width=220pt]{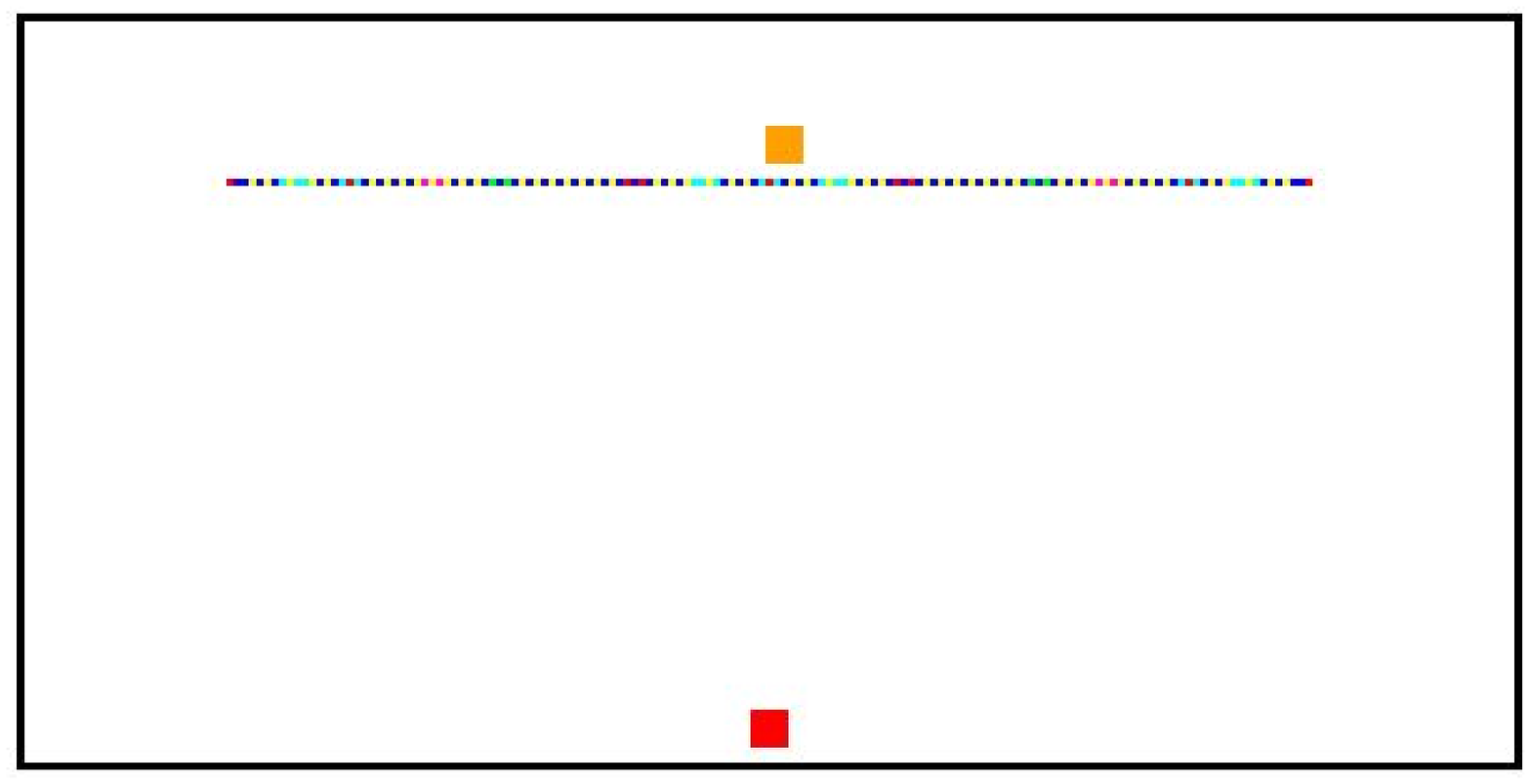} &
\includegraphics[width=220pt]{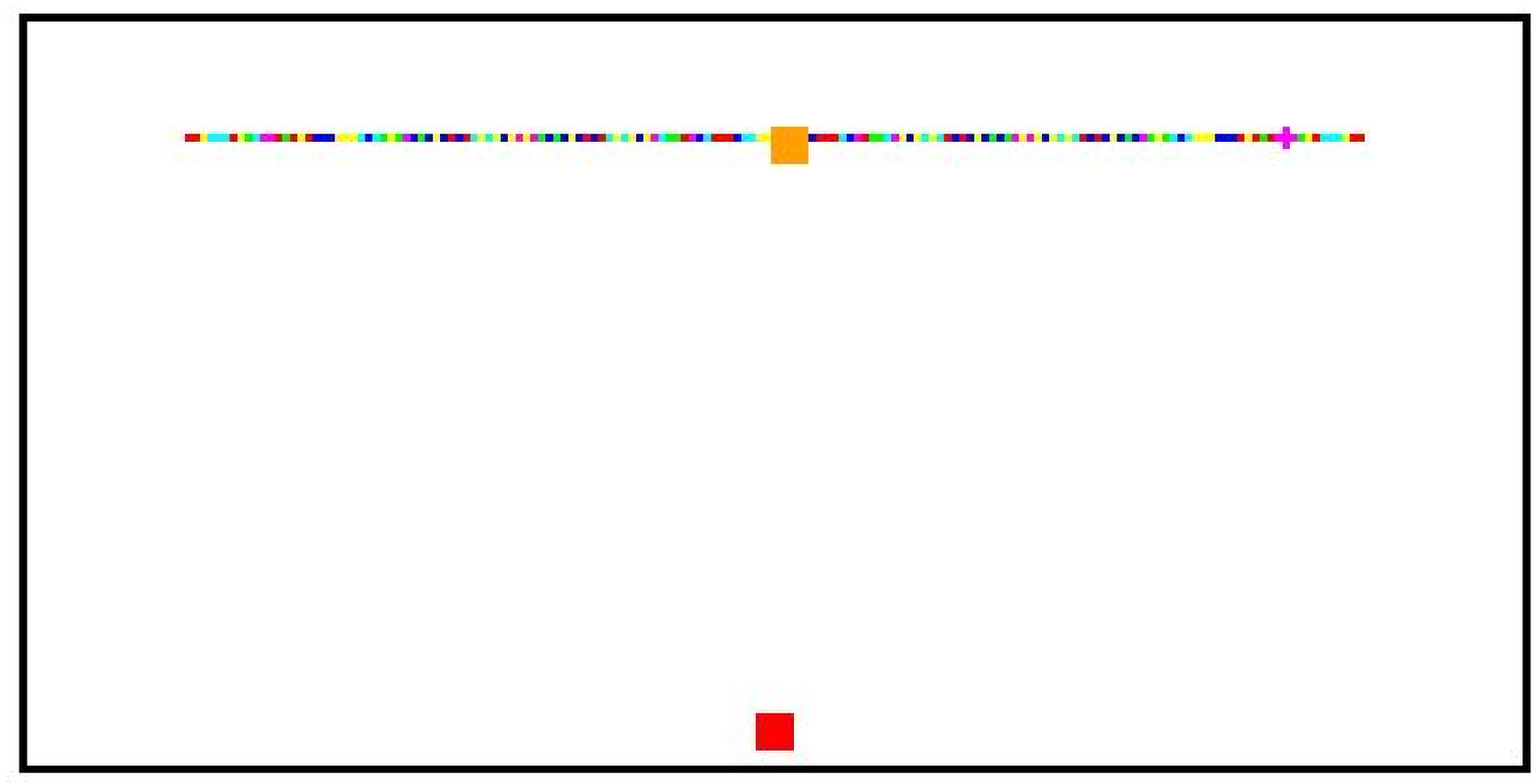} \\
\includegraphics[width=220pt]{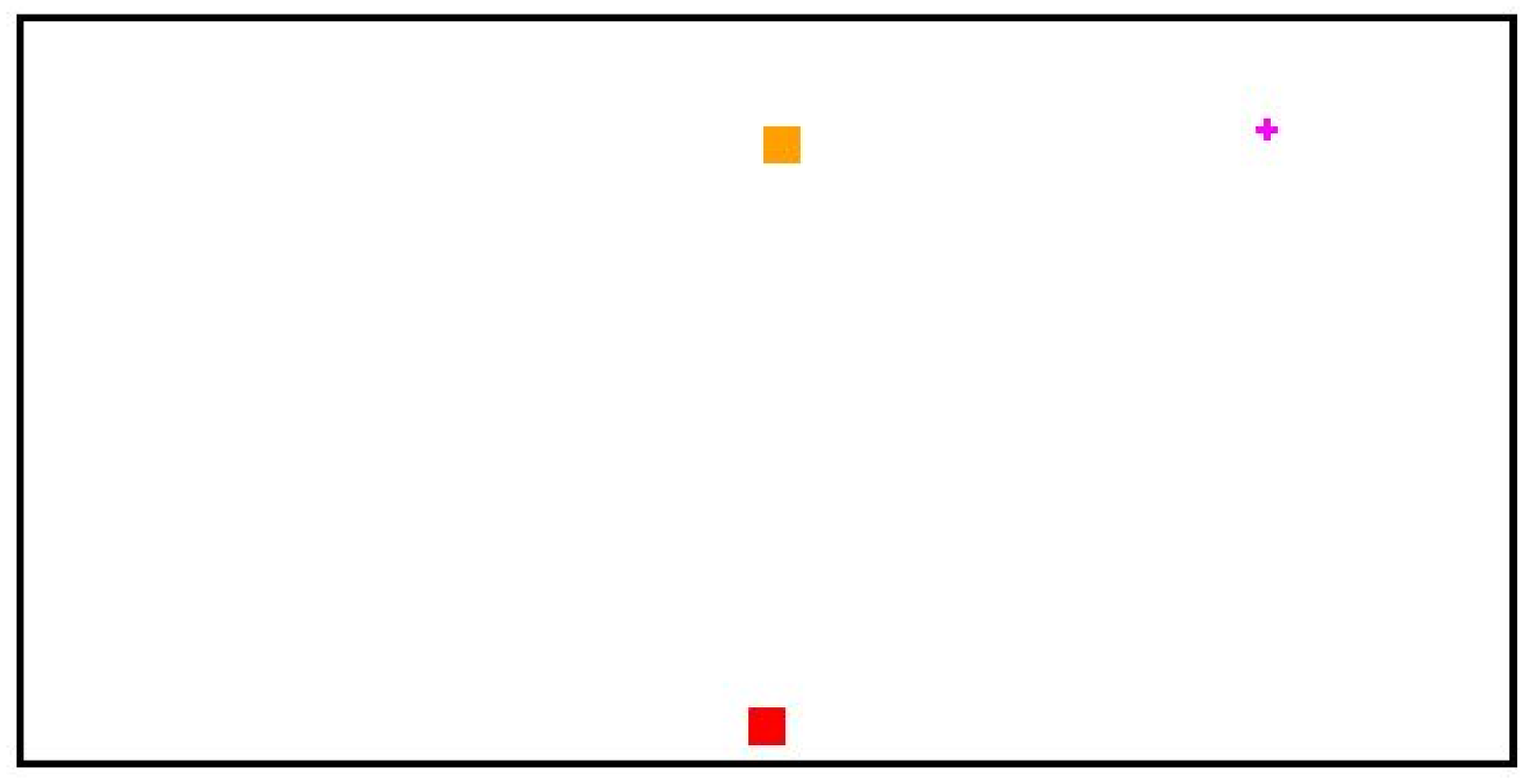} &
\includegraphics[width=220pt]{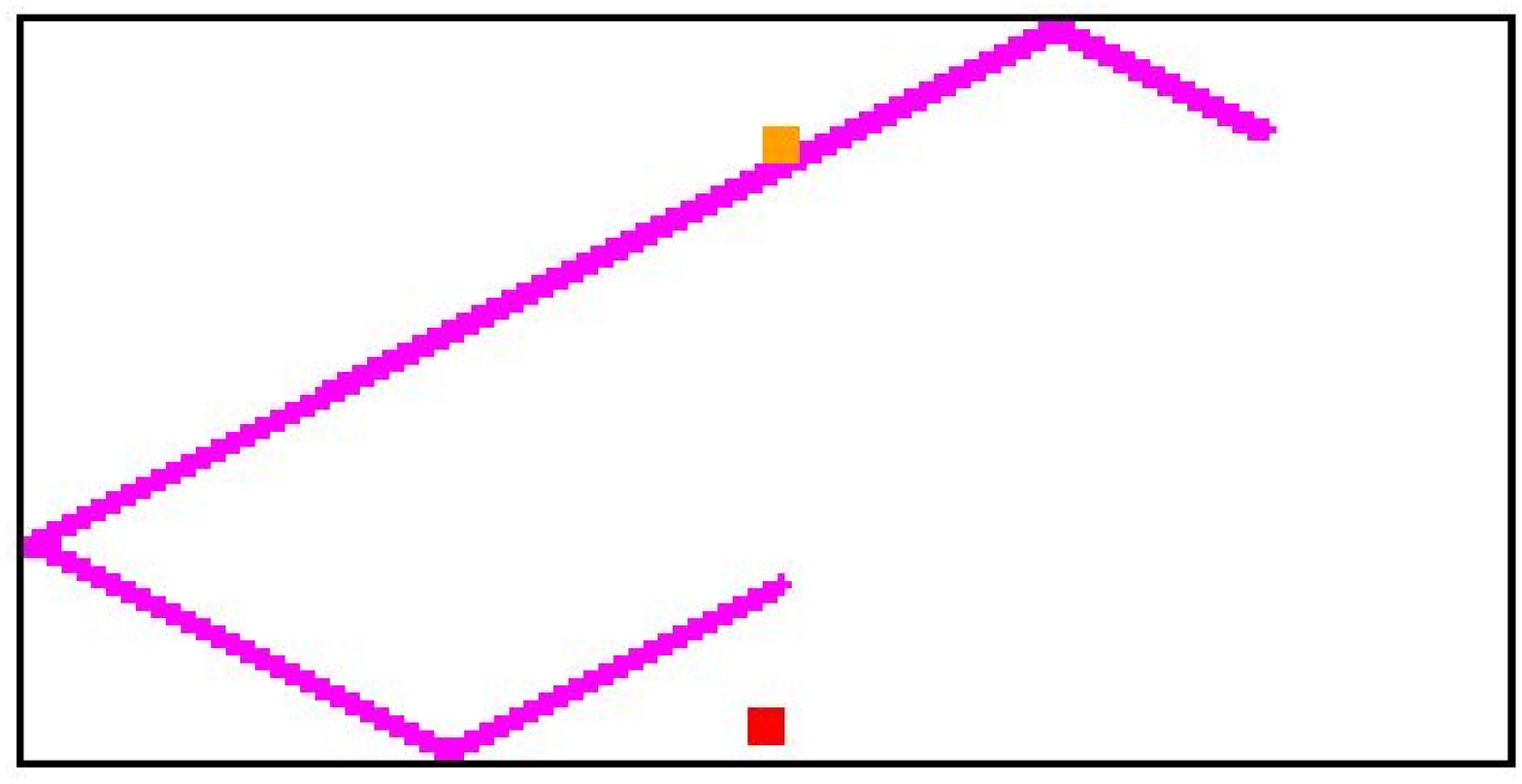} 
\end{tabular}
\end{center}
\caption{Detection Sequence}
\label{detect-sequence}
\end{figure}

Figure \ref{detect-sequence} shows a sequence of events corresponding
to the detection of a particle. The  top-left image shows the situation
just before detection. The particle is in a superposition state. The
top-right image shows the detection of the particle. The {\R} procedure is
performed and a basic state is chosen. It is actually the state of a
cell colored in pink, on the right side of the superposition state.
The bottom-left image shows the situation just after {\R}: the superposition has
disappeared and a new particle is created with the pink basic
state. The bottom-right image shows the trajectory of the particle,
bouncing on the wall, after some instants.

\begin{figure}[!htbf]
\begin{center}
\includegraphics[width=350pt]{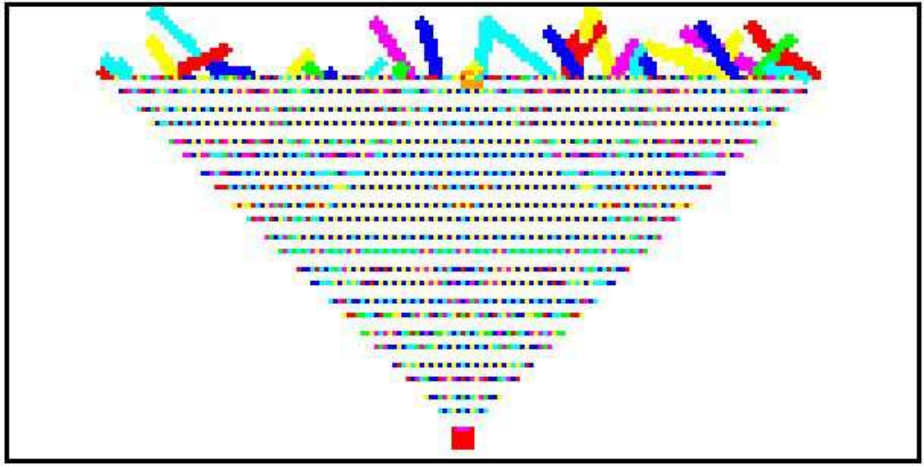}
\end{center}
\caption{Multiple Detections}
\label{mult-detection}
\end{figure}

Figure \ref{mult-detection} shows the production of a flux of
identical particles (i.e. initially in the same basic state) which are
detected in turn. Traces of the produced particles are shown on the
part of the image situated over the detector.

\subsection{Young's Experiment}

\begin{figure}[!htbf]
\begin{center}
\begin{tabular}{cc}
\includegraphics[width=220pt]{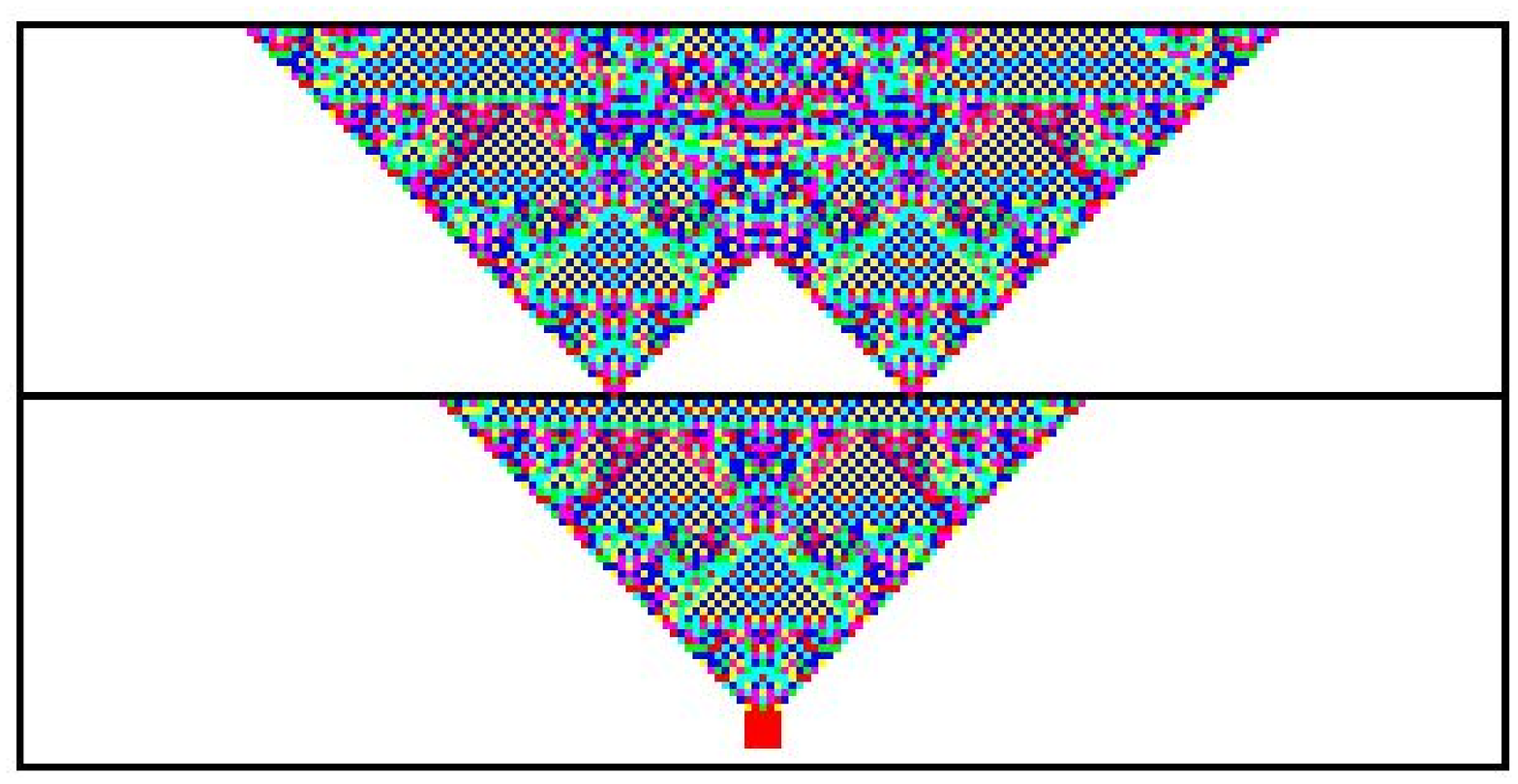} &
\includegraphics[width=220pt]{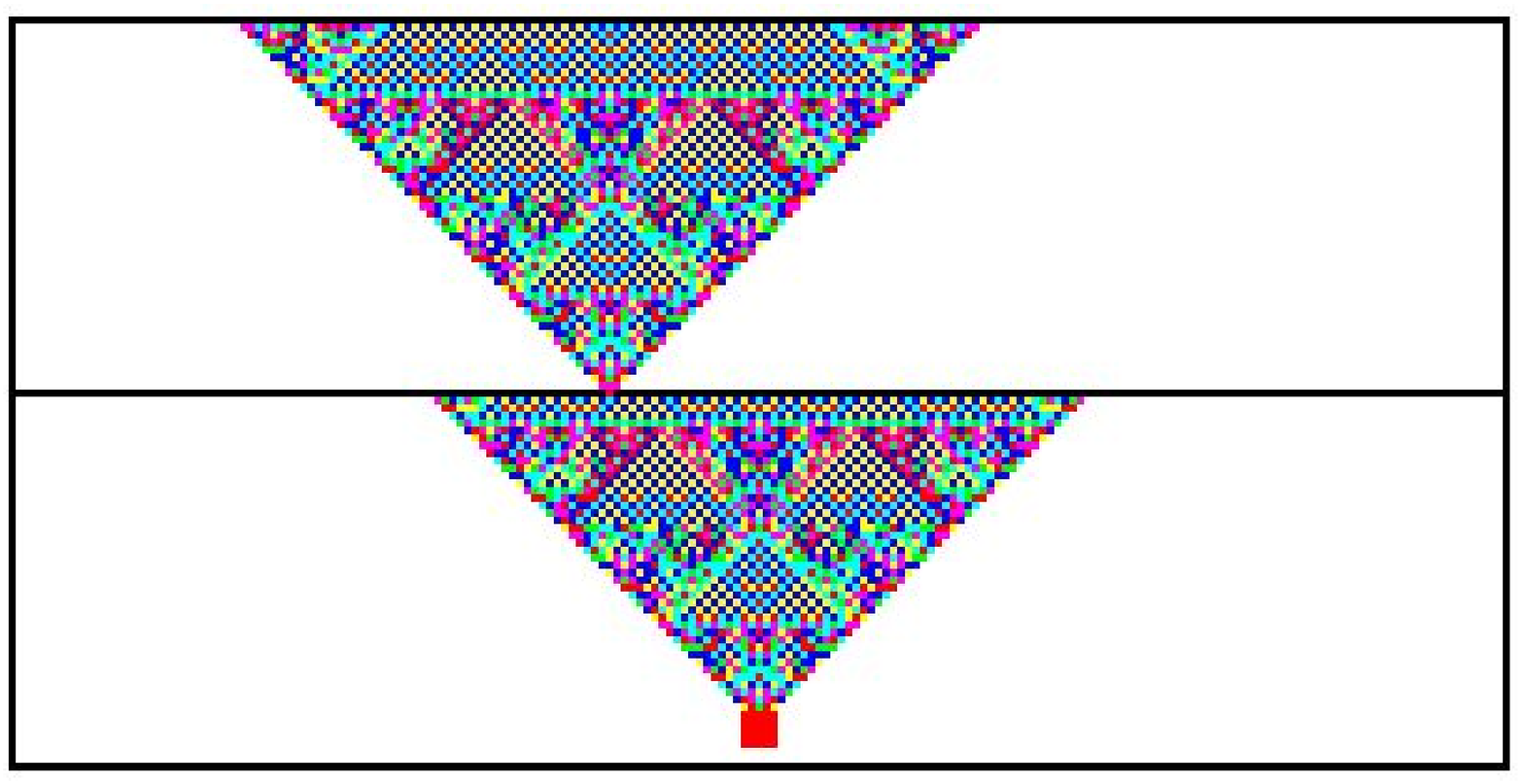} 
\end{tabular}
\end{center}
\caption{Young's Experiment}
\label{slits}
\end{figure}

A new wall is introduced in Figure \ref{slits} which separates the
simulation in two parts. In the left image, two slits are present; in
the right image, the right slit is obstructed.  This basically
corresponds to Young's experiment. In the right image, the structure
under the separation wall is reproduced over the
separation wall. In the left part, the two structures produced from the
two slits interfere when they overlap.

\begin{figure}[!htbf]
\begin{center}
\includegraphics[width=300pt]{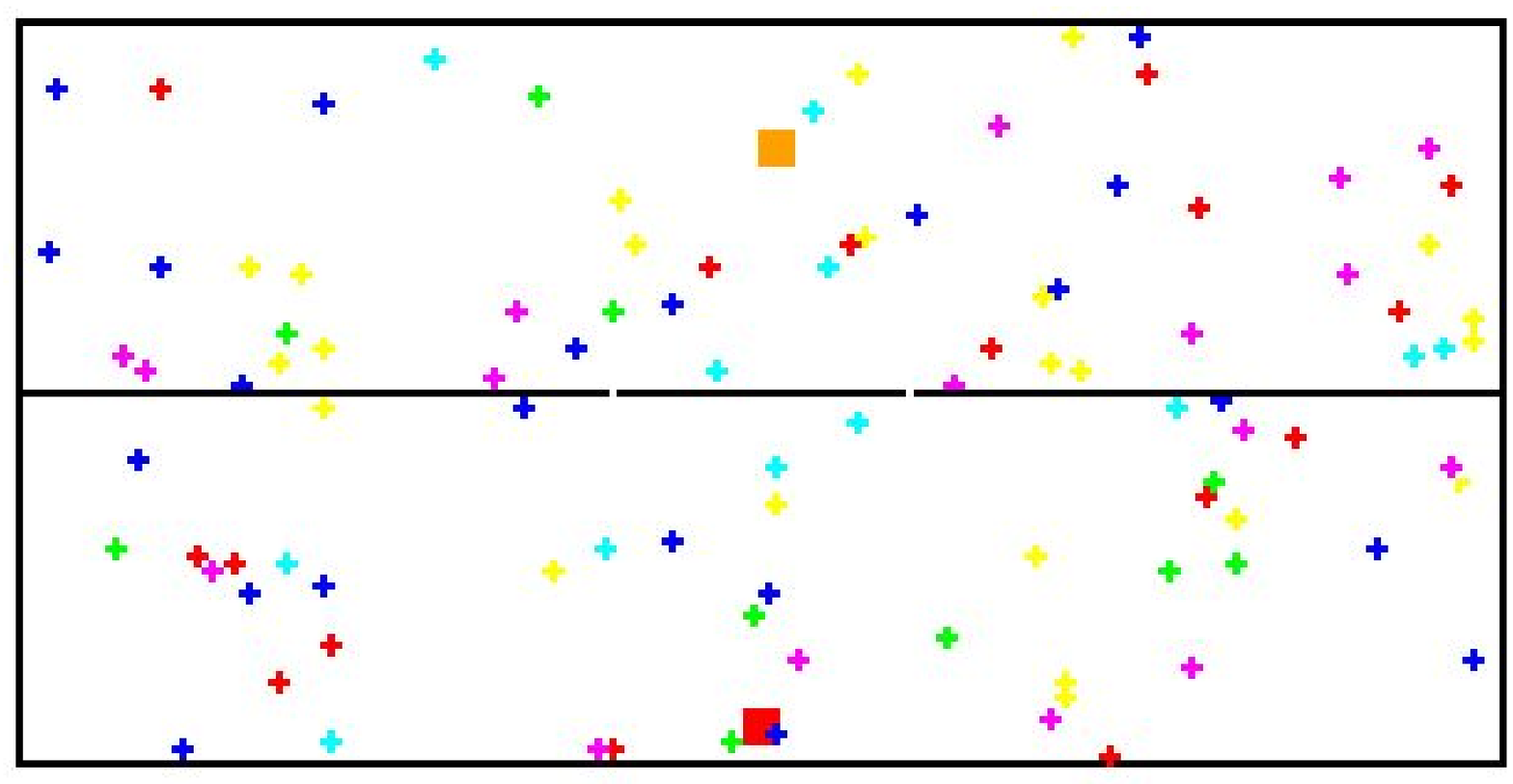}
\end{center}
\caption{Young's Experiment - 2}
\label{slits-2}
\end{figure}

We now consider the detection of particles in Young's
experiment. Figure \ref{slits-2} shows the situation after
detection of 100 particles (in presence of 2 slits).

Once the number of slits is fixed, all particles reach the detector in
the same superposition of basic states (because {\U} is deterministic and
local). When the {\R} procedure is executed, a cell is randomly chosen
from the superposition and the particle gets the cell basic state. The
probability for a particle to fall on the basic state $\ket a$ only depends
on the percentage of cells with state $\ket a$ appearing in the
superposition. The superposition state thus directly codes the
probabilities associated with the basic states. The probability of
a basic state $\ket a$ in a superposition $S$ is:

\begin{equation}
Prob (\ket a, S) = \frac {\#a \in S} {\#S}
\end{equation}

These probabilities depend on the number of slits, on the localisation
of the detector and on the localisation of the slits. Here is for
example a typical result (with 1000 particles):

\begin{center}
\begin{tabular}{|c|cccccc|}
                                                                                 \hline
basic state      &0     &1    &2    &3    &4    &5      \\ \hline
1 slit               & 0.327 & 0.023 & 0.281 & 0.153 & 0.017 & 0.199   \\ \hline
2 slits             & 0.226 & 0.039 & 0.298 & 0.171 & 0.116 & 0.150    \\ \hline

\end{tabular}
\end{center}
\noindent
In the configuration from which this table is produced, a particle
has for example probability 0.327 to fall in state 0 with 1 slit,
while it has probability 0.226 to fall in the same state with 2 slits.


\subsection {Entanglement}

\begin{figure}[!htbf]
\begin{center}
\begin{tabular}{cc}
\includegraphics[width=220pt]{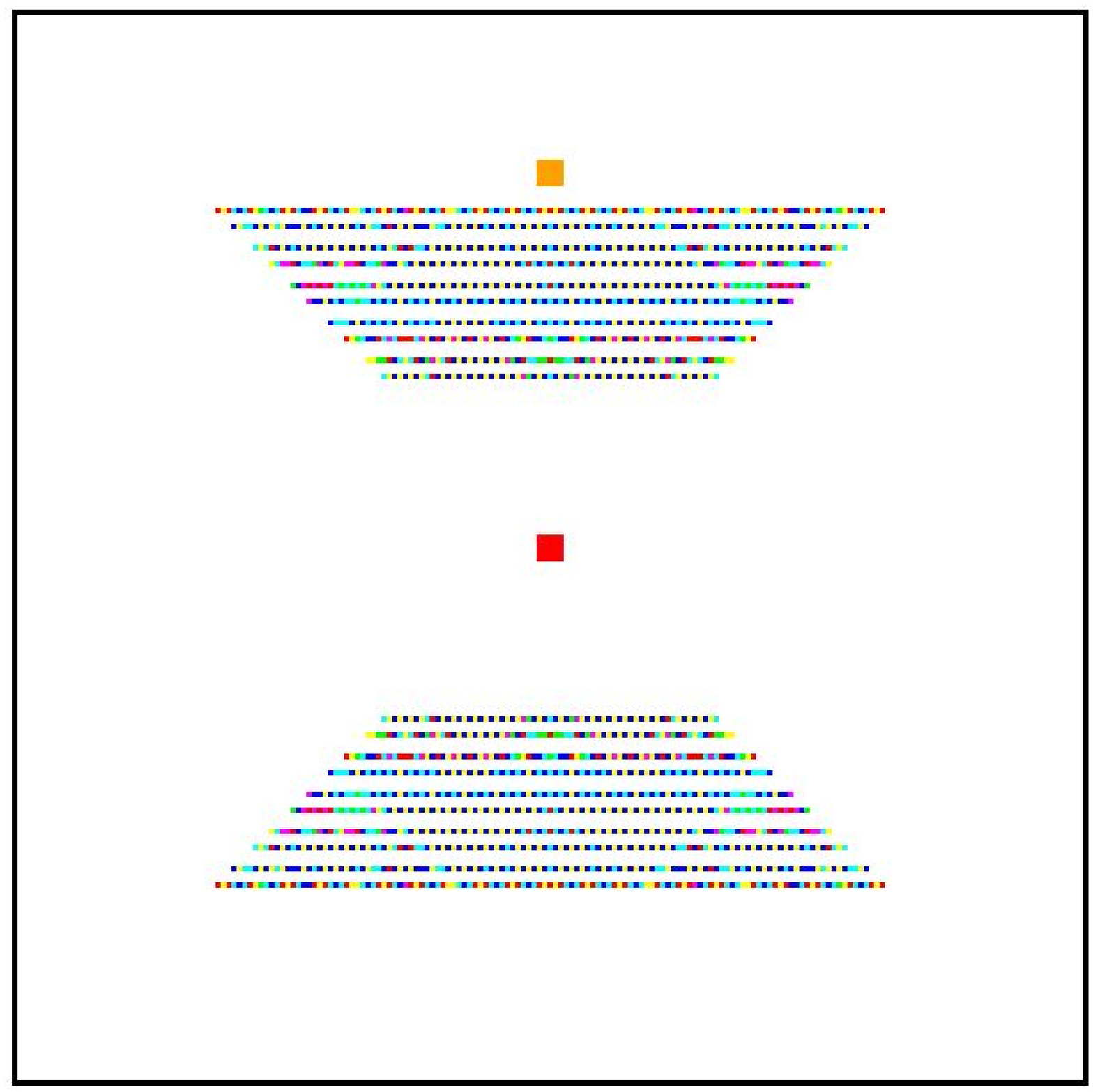} &
\includegraphics[width=220pt]{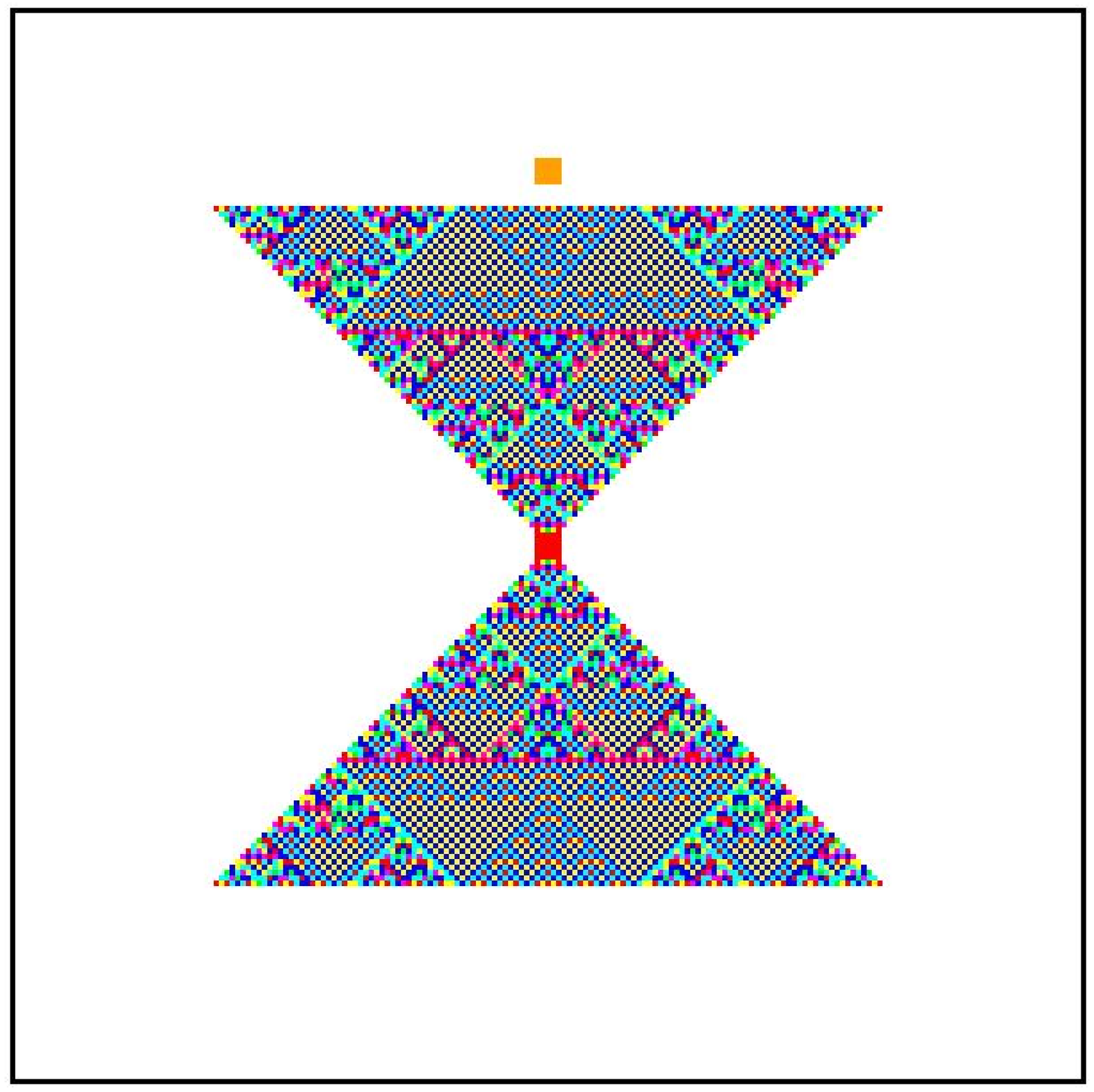} \\
\includegraphics[width=220pt]{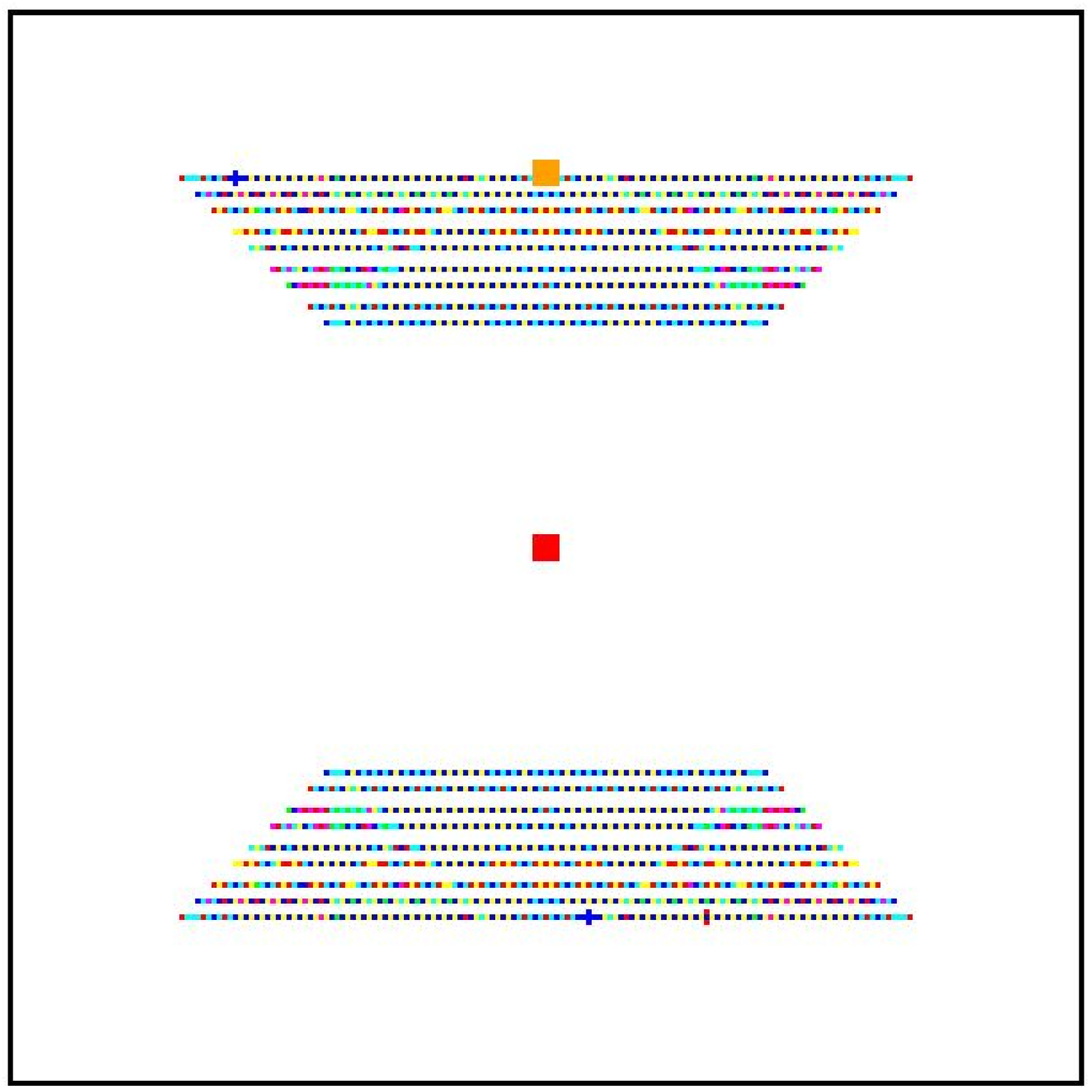} &
\includegraphics[width=220pt]{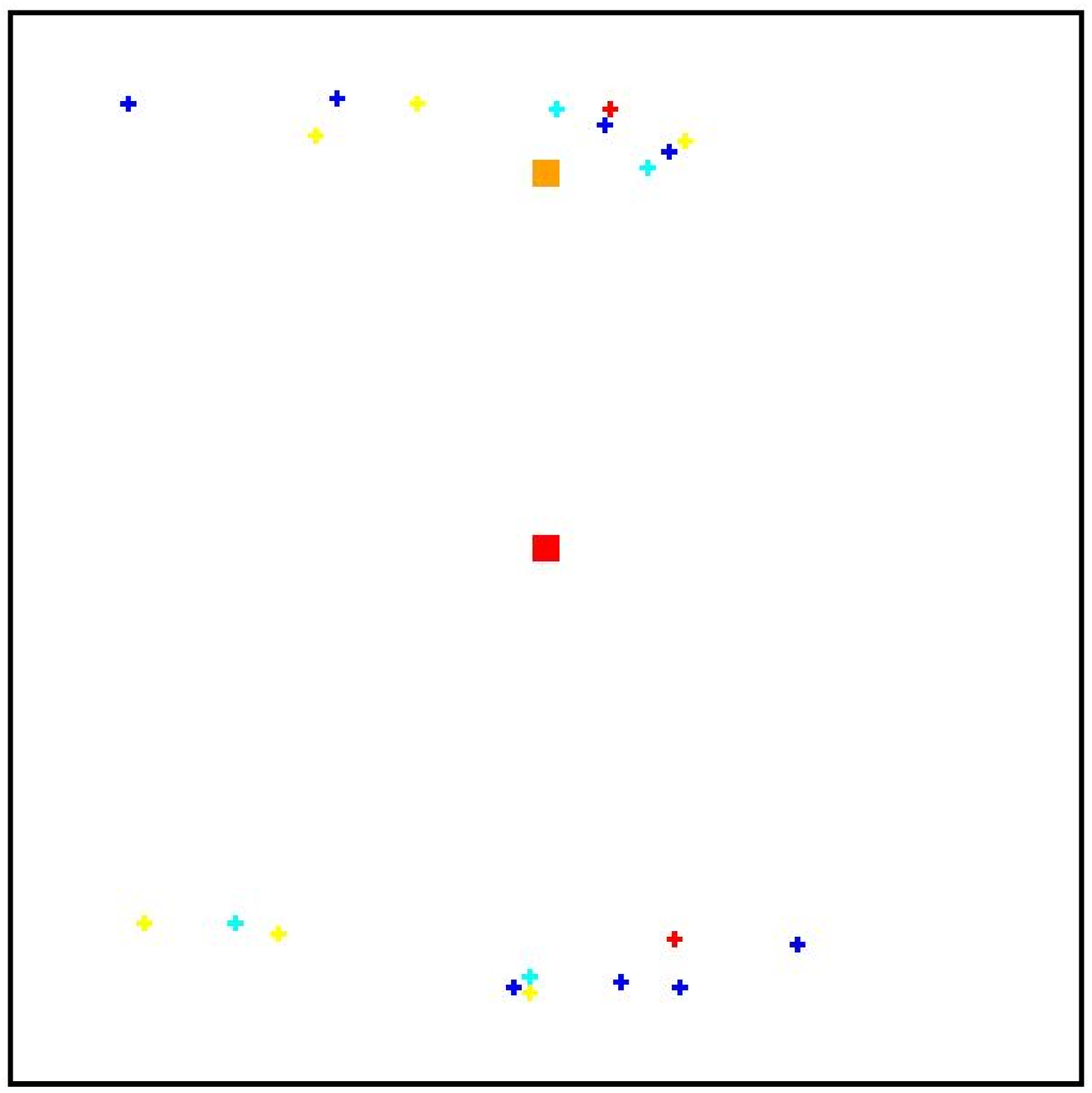} 
\end{tabular}
\end{center}
\caption{Entanglement Sequence}
\label{entangl-sequence}
\end{figure}

We now consider pairs of entangled particles emitted by the same
source. Each pair is composed of a particle emitted in the top direction
and an entangled particle emitted in the bottom direction.
Entanglement basically means two things: first, the two particles share
the same {\R} procedure; second, the choices of basic states implied by {\R}
are not independent (to simplify, we consider that the two particles 
always choose the same basic state).  An
example of 10 emissions of pairs of entangled particles is shown in
Figure \ref{entangl-sequence}:

\begin{itemize}
\item The   top-left image shows the situation where all the pairs have been emitted by the
  source but none has yet been detected. Two beams of particles are
  present, one directed to the top and one to the bottom.

\item The same situation is shown on the  top-right image, but 
  traces of the particles are kept visible (remanent images).

\item The bottom-left image shows the detection of a particle (blue) by
  the detector (big orange square). The entangled particle (also blue)
  appears at the bottom of the image.

\item In the bottom-right image, all 10 particles have been detected. The
10 corresponding entangled particles are present at the bottom of the image.
\end{itemize}

\section {Implementation \label {section:implementation}}

The simulation is implemented in the FunLoft programming language
\cite{Boussinot-FunLoft,indes-FunLoft}. CA's and particles have been
previously implemented in several contexts of reactive programming
(e.g. in the two research papers \cite{Hal-Boussinot-ca} and
\cite{FBFunLoftBench07}; the simulation described here uses elements
of both papers).

\begin{figure}[!htbf]
\begin{center}
\begin{tabular}{c|cc}
instant $t$ & \multicolumn{2}{c}{instant $t+1$}  \\ \hline
collect & increment & transmit \\
\includegraphics[width=70pt]{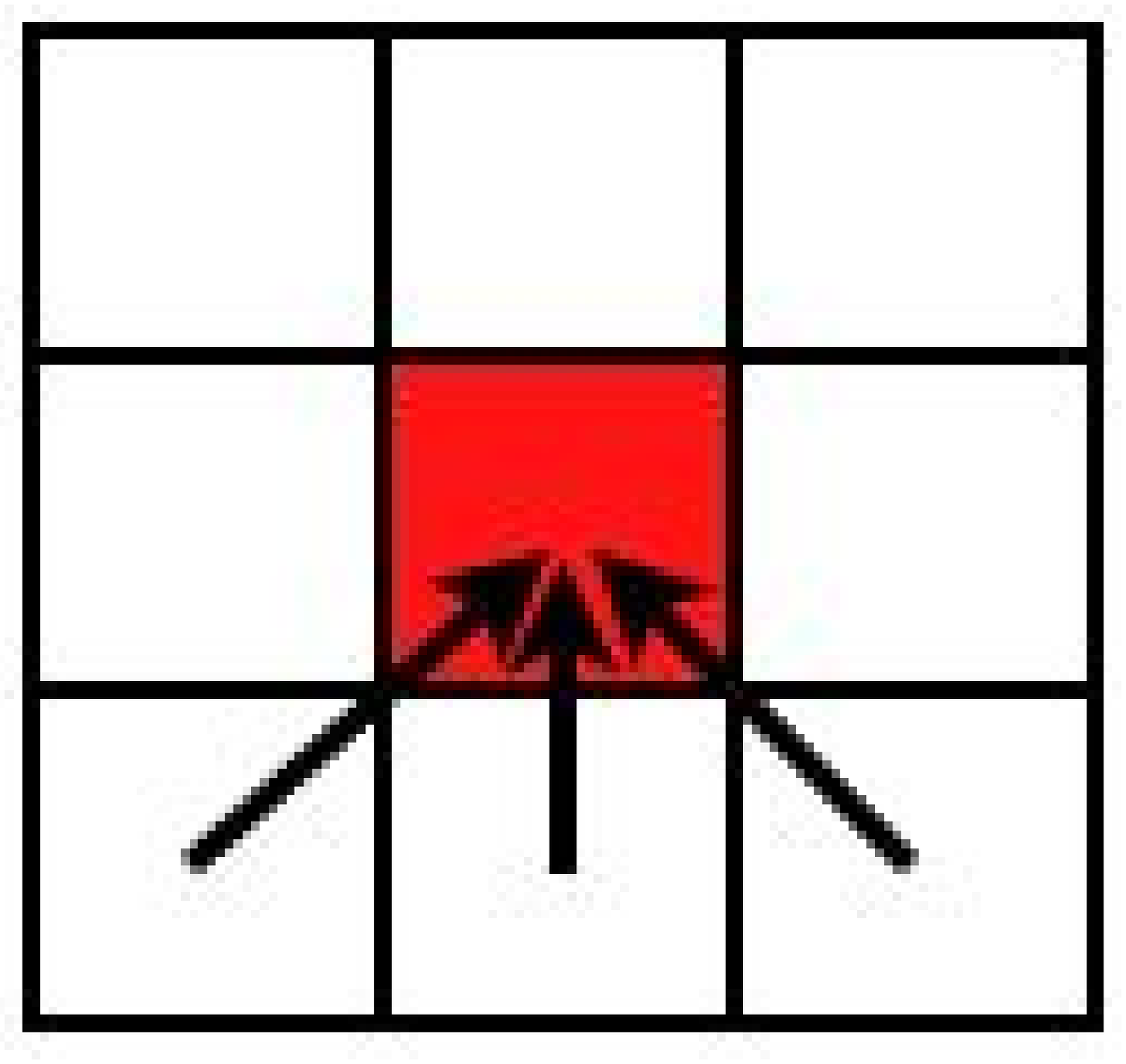} &
\includegraphics[width=70pt]{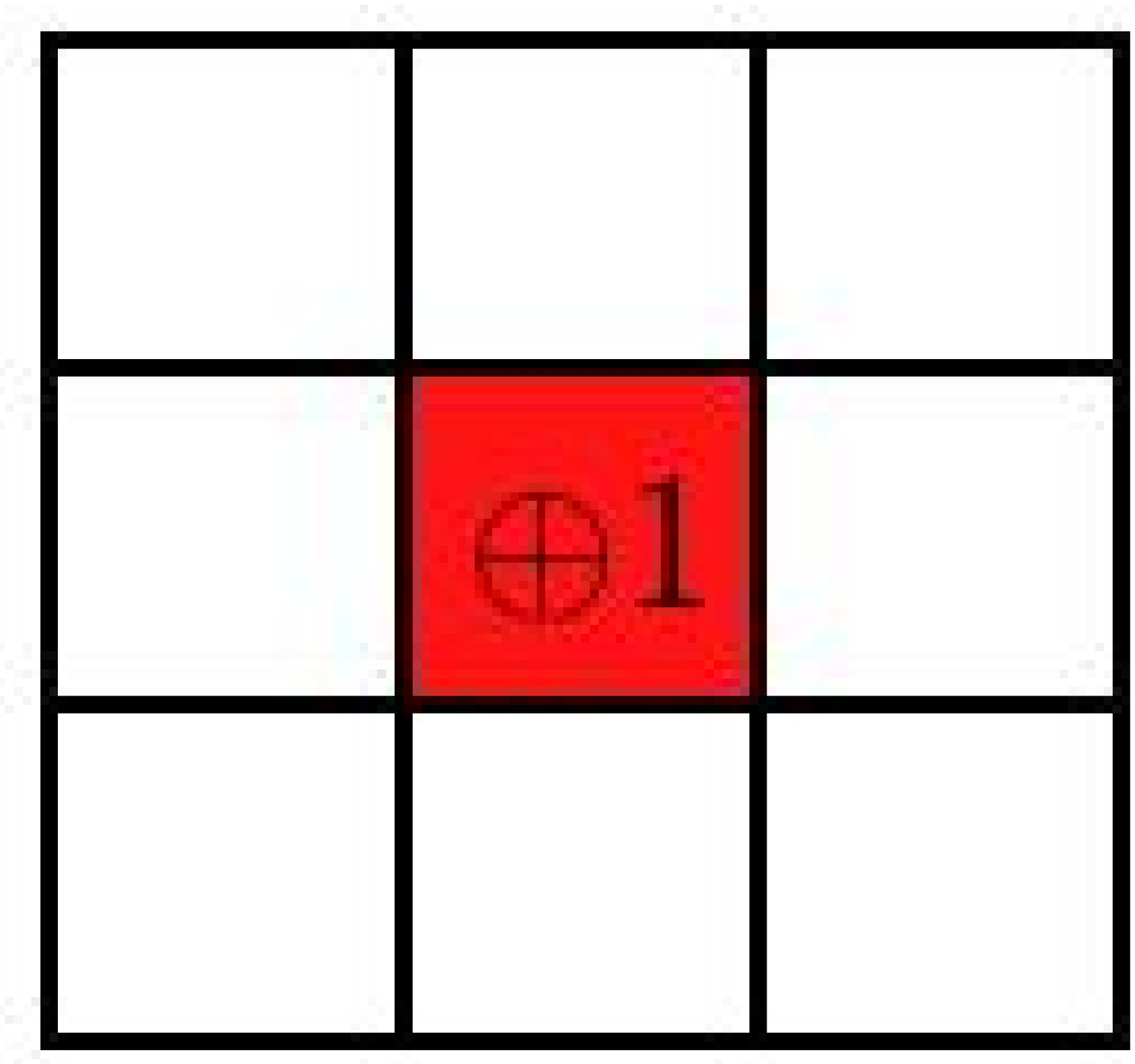} &
\includegraphics[width=70pt]{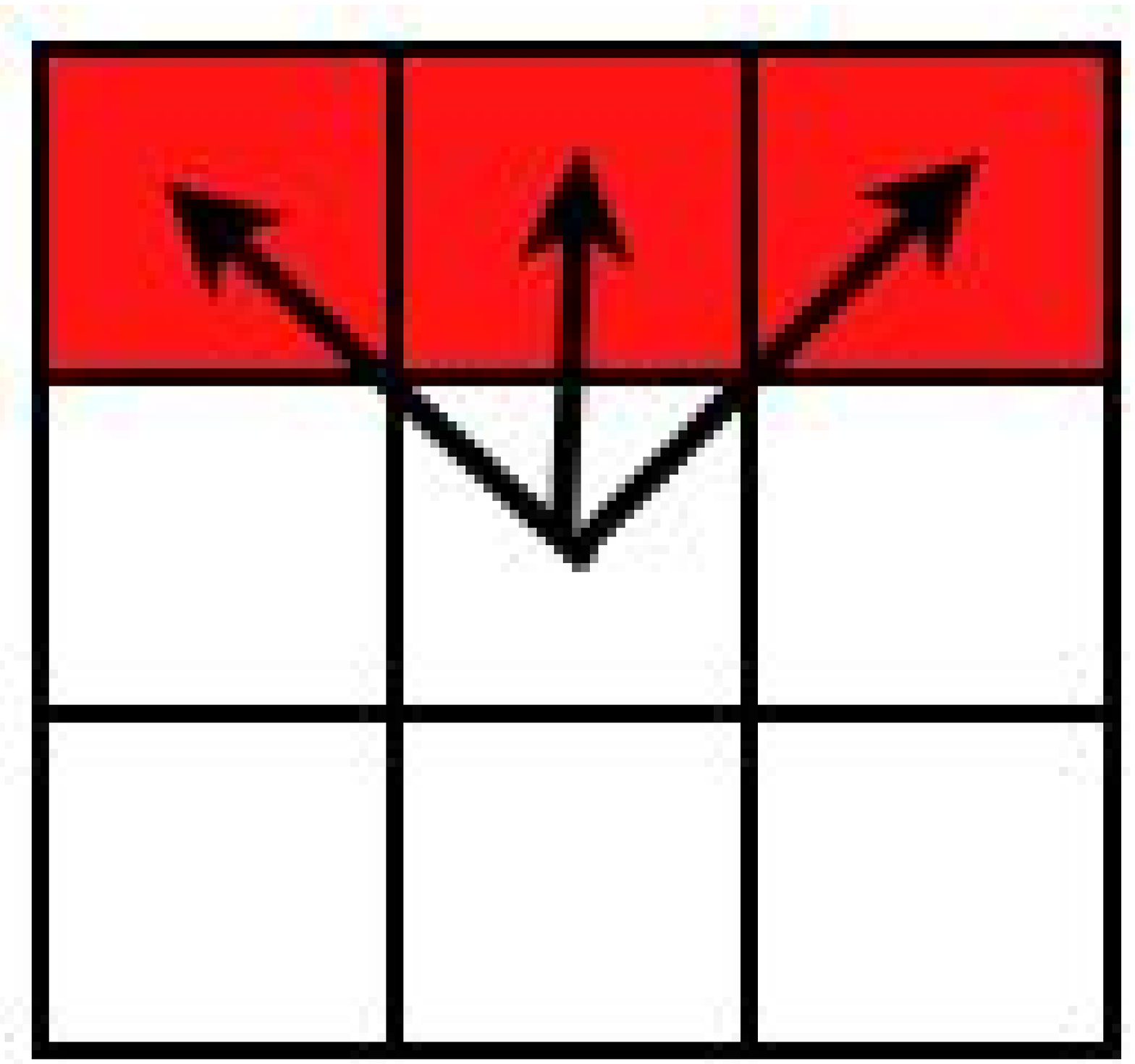} \\
$s_y^x  \assign s_{y+1}^{x-1}  \plusmod s_{y+1}^{x}  \plusmod s_{y+1}^{x+1}$ &
$s_y^x  \assign s_y^x \plusmod 1$& 

\end{tabular}
\end{center}
\caption{Cell Rule}
\label{cell-rule}
\end{figure}

Basically, we consider a cellular automaton embedded in a reactive
simulation. The reactive simulation runs reactive threads (simply
called threads in the sequel) in a cooperative way, and defines global
instants shared by all the threads. According to the presence of
instants, threads can synchronise and communicate with broadcast
events. Instants and evolution steps of the cellular automaton are
identified. Each cell of the underlying cellular automaton is
implemented by a thread; moreover, particles created in response to
measures are also implemented with threads. The cell rule is described
in Figure \ref{cell-rule} (for the top direction). A cell
living at instant $t$ (in red) collects the activations of the 3 cells
under it and increments its state by the collected values
($\plusmod$ denotes addition modulo 6); then, at instant $t+1$, the
cell increments its state by 1, and transmits it to the 3 cells above it.

Cells that are living at instant $t$ form the superposition
state at instant $t$.  When a measure occurs, an event is broadcast to
all the living cells, which stops their transmission step. Moreover,
one of the living cells is randomly chosen and a particle is created
from it.

The simulation is approximatively made of 1400 lines of code; in this
section, we only give the main parts of the code. The section is
structured as follows: the data structure of cells is described in Section \ref
{implem:cell}. Section \ref {implem:neighbours} describes the
activation of cells ({\U} procedure). The cell behaviour combining
both {\U} and {\R} is described
in \ref {implem:behav}. Section \ref {implem:reduce} describes the
{\R} procedure. The source of single particles is described in \ref
{implem:source} and the source of entangled particles is described in
\ref {implem:entanglement}.

\subsection {Cell Data Structure \label {implem:cell}}

The data structure associated with cells is the following:

{\small
\begin{verbatim}
type cell_t = cell of 
    x: int
  * y: int
  * kind: cell_kind_t ref
  * living: bool ref
  * basic_state: int ref
  * trigger: (activation_t) event_t 
  * R: (unit) event_t ref
  * signal: (int) event_t ref
  * chosen: int ref ref
  * chosen_state: int ref ref
\end{verbatim}}

\noindent
In this definition:
\begin{itemize}
\item \code{x} and \code{y} are the cell coordinates;
\item \code{kind} is the cell kind, that is, the direction of the
  particle (\code{UP} or \code{DOWN}), or \code{BRICK} for cells belonging to a wall; 
\item \code{living} holds true when the cell is living and false when it is dead;
\item \code{basic\_state} holds the basic state associated with the cell;
\item \code{trigger} holds the event used by neighbours to activate
  the cell; the event has a value of type \code{activation\_t} which
  describes the characteristics of the triggering neighbour (see \ref{implem:neighbours});
\item \code{R} holds the event used by the {\R} procedure;
\item \code{signal} holds the event used to signal the basic state of the cell;
\item \code{chosen} holds the chosen cell, when chosen, and -1 otherwise.
\item \code{chosen\_state} holds the chosen state, when chosen, and -1 otherwise.
\end{itemize}

Note that \code{chosen} and \code{chosen\_state} are doubly referenced
because they can be both communicated and changed (as example,
\code{chosen\_state} is needed for entanglement, in order to be be
shared among the entangled particles).

Basic states are integers modulo \code{base}; we define \code{add\_state} that adds an integer to a cell basic state, and
\code{increm\_state} that increments a cell basic state:

{\small
\begin{verbatim}
let add_state (c,z) = 
  let a = cell.basic_state (c) in
    a := (!a + z) mod base

let increm_state (c) = add_state (c,1)
\end{verbatim}}

\subsection {Neighbours \label {implem:neighbours}}

Cells trigger their neighbours by giving them an {\it activation}
record which is a data-structure defined by:

{\small
\begin{verbatim}
type activation_t = activation of
  * kind: cell_kind_t
  * basic_state: int
  * R: (unit) event_t
  * signal: (int) event_t
  * chosen: int ref
  * chosen_state: int ref
\end{verbatim}}

In order to trigger a neighbour, a cell generates the trigger event of the neighbour 
with an activation record describing  itself (nothing is done if the neighbour is a brick):

{\small
\begin{verbatim}
let awake_neighbour (c,ix,iy) =
  let neighbour = !cell_array [cell.x (c) + ix,cell.y (c) + iy] in
  let state = !cell.basic_state (c) in
  let kind = !cell.kind (c) in
  let r = !cell.R (c) in
  let signal = !cell.signal (c) in
  let chosen = !cell.chosen (c) in
  let chosen_state = !cell.chosen_state (c) in
    if !cell.kind (neighbour) <> BRICK then
      generate cell.trigger (neighbour) with 
        activation (kind,state,r,signal,chosen,chosen_state)
    end
\end{verbatim}}

Here is the part corresponding to the \code{UP} direction of the function that triggers the neighbourhood of a cell:

{\small
\begin{verbatim}
let awake_neighbourhood (c) =
  let k = !cell.kind (c) in
    if k = UP then
      begin
        awake_neighbour (c,-1,-1);
        awake_neighbour (c,0,-1);
        awake_neighbour (c,1,-1);
      end
    else 
      ...
\end{verbatim}}
  The cells triggered are the cell \code{c1} situated immediately above \code{c},
  and the two cells at the right and at the left of \code{c1}.

  The other directions are processed in a similar way that we do not
  describe here for the sake of conciseness.

  When triggered by a neighbour, a cell combines the associated
  activation record with itself. We choose a very simple combination
  scheme: the state of the triggering cell is added to the state of
  the activated cell. The function \code{combine} is defined by:

{\small
\begin{verbatim}
let combine (c,activ) =
  begin
    cell.kind (c) := activation.kind (activ);
    add_state (c,activation.basic_state (activ));
    cell.R (c) := activation.R (activ);
    cell.signal (c) := activation.signal (activ);
    cell.chosen (c) := activation.chosen (activ);
    cell.chosen_state (c) := activation.chosen_state (activ);
  end
\end{verbatim}}

\subsection {Cell Behaviour \label {implem:behav}}

The {\U} procedure basically implements the behaviour described
in Figure \ref{cell-rule}: each cell
living at instant $t$ collects the activations from the
cells below it, adding their states to its current state; then, at instant
$t+1$, the cell increments its state and activates the
three cells above it. A cell thus transmits to the cells above the
collection of the states of the cells below it, after having
incremented its own state.  More precisely, the behaviour of a cell is
the following (we do not consider here the case of
bricks)\footnote{comments start by \code{//} and continue up to the end of line.}:

{\small
\begin{verbatim}
let module cell_behavior (c) = 
  loop
    begin 
      // step1
      if !cell.living (c) then awake_neighbourhood (c)
      else 
        begin 
          await cell.trigger (c); 
          cell.living (c) := true; 
        end;
      // step2
      for_all_values cell.trigger (c) with a -> combine (c,a);
      increm_state (c);
      display (c);
      // step3
      let r = ref Nil_list in
        begin
          get_all_values !cell.R (c) in r;
          if length (!r) <> 0 then // R called
            let done = event in
              begin
                thread reduce (c,done);
                await done;
              end
          else 
            awake_neighbourhood (c)
        end;
      // step4
      cell_reset (c);
    end
\end{verbatim}}

\noindent
This cyclic behaviour (\code{loop}) is composed of the following steps:

\begin{itemize}
\item [step1:] if the cell is living, then the neighbours are triggered (\code{awake\_neighborhood}). Otherwise, the cell awaits to be triggered (\code{await}).

\item [step2:] the cell collects all the values generated by its neighbours (\code{for\_all\_values}) and combines them (\code{combine}). Then, the cell increments its state (\code{increm\_state}) and displays itself on screen (\code{display}).

\item [step3:] the possibility of reduction with the {\R} procedure is
  considered. All the values associated with the \code{R} event are
  collected (\code{get\_all\_values}). If the collected list is not
  empty, then an instance of the \code{reduce} module is launched and
  its termination is awaited using the new event \code{done}.
Otherwise, the neighbours are triggered.

\item [step4:] the cell is reset (\code{cell\_reset}); this function
  basically resets the basic state, records the cell as dead, and
  erases it on screen.  

\end{itemize}

\subsection {Reduce Procedure \label {implem:reduce}}

Let us consider the module that chooses a basic state from a
superposition. First, it awaits the signal event which is generated by
all the cells involved in the superposition when the {\R} procedure is
executed. Each cells gives its identity (under the form of a linear
combination of its coordinates) together with the signal. Then, all
the identities are collected and one is chosen nondeterministically
using the variable \code{chosen} (initially, \code{chosen} equals -1;
when the first choice occurs, \code{chosen} is set to the identity of
the chosen cell, returned by the function \code{choose}):

{\small
\begin{verbatim} 
let module choose_in_superposition (c) =
  let signal = !cell.signal (c) in
  let chosen = !cell.chosen (c) in
  let collect = ref Nil_list in
    begin
      await signal;
      get_all_values signal in collect;
      if !chosen = -1 then 
        chosen := choose (!collect) 
      end
    end
\end{verbatim}}

The module that implements the {\R} procedure first launches a thread to
choose a basic state. At the next instant (to let the launched thread
start execution), it signals itself. Then, after two
instants, if it is chosen then it sets the chosen state (\code{set\_chosen\_state}) and
launches a particle bearing the color of the chosen state (\code
{particle\_behavior}). The code is:

{\small
\begin{verbatim}
let set_chosen_state (c) =
  if !!cell.chosen_state (c) = -1 then
     !cell.chosen_state (c) := !cell.basic_state (c)
  end

let module reduce (c,done) =
  let x = cell.x (c) in
  let y = cell.y (c) in
  let me = linear (x,y) in
    begin
      thread choose_in_superposition (c);
      cooperate;
      generate !cell.signal (c) with me;
      cooperate;
      cooperate;
      if !!cell.chosen (c) = me then
        begin 
          set_chosen_state (c);
          let fx = int2float (x) in
          let fy = int2float (y) in
          let dir = cell.kind (c) in
          let state = !!cell.chosen_state (c) in
            thread particle_behavior (fx,fy,cell_color (state),dir)
        end 
      end;
      generate done
    end
\end{verbatim}}

\noindent
Note that the function \code{choose} is actually the only source of randomness in the simulation.

\subsection {Source \label {implem:source}}

The firing function generates the triggering event of a cell, with an
activation made of a new signal and a new reference holder for
\code{chosen}:

{\small
\begin{verbatim}
let fire (c,a,dir,e,s) =
  generate cell.trigger (c) with activation (a,e,dir,event,ref -1,s)
\end{verbatim}}

  A (standard) source cyclically fires a cell when receiving the event
  \code{go}. Each fired particle has a new event \code{R} and a new
 state holder for \code{chosen\_state}. All particles are initially
  in the same basic state.

{\small
\begin{verbatim}
let module source_r (x,y,s,dir,go) =
  let c = !cell_array [x,y] in
    loop
      begin
        await go; 
        fire (in_dir (c,dir),s,dir,event,ref -1);
        cooperate;
      end
\end{verbatim}}

\subsection {Entanglement \label {implem:entanglement}}

A source of entangled particles emits two particles in opposite
directions, but sharing the same \code{R} event and the same state
holder. Thus, when one of the entangled particles is detected, it
is exactly as if the other was also detected, at the same instant;
moreover, the choice of the basic state made for the detected particle
is instantaneously transmitted to the other particle. The code of the
source producing entangled particles in opposite directions is:

{\small
\begin{verbatim}
let module dual_source_r (x,y,s,dir,inv,go) =
  let c = !cell_array [x,y] in
    loop
      let r = event in    // shared R
      let sh = ref -1 in  // shared state holder
        begin
          await go;
          fire (in_dir (c,dir),s,dir,r,sh);
          fire (in_dir (c,inv),s,inv,r,sh);
          cooperate;
        end
\end{verbatim}}

Note that the state holder is a ``shared reference'' which can be
assigned and read by entangled particles. These shared references are
different from ``hidden variables'' that would be set at creation of
the entangled pair, but that would not be readable (``hidden'') by any
measure. In the simulation, there is actually an instantaneous
(i.e. during the same instant) communication that occurs between the two entangled
particles when {\R} occurs (and not before that moment).


\section {Related Work \label {section:related-work}}

Simulation of physics with computers has been initiated by R. Feynman in \cite{Feynman:1982}.
His paper ends with:
{\it ``And I'm not happy with all the analyses that go with just the classical theory, because nature isn't classical, dammit, and if you want to make a simulation of nature, you'd better make it quantum mechanical, and by golly it's a wonderful problem, because it doesn't look so easy.''}
Feynman directly spoke to computer scientists, saying:
{\it ``... what I'm trying to do is to get you people who think about computer-simulation possibilities to pay a great deal of attention to this, to digest as well as possible the real answers of quantum mechanics, and see if you can't invent a different point of view than the physicists have had to invent to describe this.''} The initial motivation for the simulation proposed here was to understand quantum mechanics. And to understand something, what's could be better than to implement it?

In \cite{Hayes:2003}, B. Hayes discusses the issue of writing programs to implement a physically plausible universe. He considers the use of cellular automata and notes that {\it ``The global clock is surely an uninvited guest in the world of cellular automata. Having taken pain to create a purely local computational physics, we then introduce a signal that must be broadcast to every point in the universe in every moment of time. What could be} less local {\it than that?''}. This remark is totally relevant for the simulation presented here, but we don't see any alternative to model instantaneous aspects of QM (as a matter of fact, we see instantaneity as the strangest aspect of QM).

In \cite{Wheeler:1982} J. A. Wheeler compares nature and computers. He says {\it ``Time is not a primary category in the description of nature. It is secondary, approximate, and derived.''} Our simulation actually introduces two levels of time. The first level is the level of instants which is global to both CA cells and (real) animated particles; at that level, events are completely ordered and causality is totally defined.  The second level concerns the events occurring during the same instant; they are not totally ordered and the causality is a ``loose'' one in which the notions of ``before'' and ``after'' are not totally strict and depend on the implementation.

To consider the universe as a computing machine ({\it Digital Physics}) is an idea which may be found for example in K. Zuse's work, where the computer is a cellular automaton. S. Lloys \cite {Lloyds:2006} considers a quantum computer instead. Here, we describe a cellular automaton embedded in a reactive environment, naturally introducing global instants and broadcast events, and we do not consider any aspect of quantum computing.


\section{Conclusion\label{section:conclusion}}

We have implemented a simulation with the following characteristics:
\begin{itemize}

\item The simulation is composed of a cellular automaton embedded in a reactive environment providing global instants and broadcast events.  

\item The basic components are threads that are executed in synchronous parallelism. Each cell of the cellular automaton is implemented by a thread. After detection, each particle is animated by several threads of the reactive environment.

\item The cells living at instant $t$ code the superposition state of the particle at instant $t$. Before detection, the evolution of the particle is defined by the cellular automaton rule ({\U} procedure).

\item Detection of a particle generates a broadcast event, instantaneously received by all the corresponding living cells of the cellular automaton.

\item After detection, living cells are instantaneously reset, and a particle animated by the reactive environment is created. The state of the created particle is randomly chosen among the basic states composing the superposition  ({\R} procedure). This is the only source of nondeterminism in the simulation.

\item Entanglement of particles basically means that they share the same procedure {\R}. The shared event corresponding to {\R} is instantaneously broadcast to all entangled particles by the reactive environment.

\end{itemize}

There are at least two regards in which the simulation could certainly be improved:
\begin{itemize}
\item In the current simulation, superposition states are segments; they could be replaced by circular shapes centered on the source. Hexagonal CA's could be useful for that purpose.

\item One may wonder how to introduce gravity, for particles that have a mass.  With real particles, this is not a problem: gravity can be naturally implemented as a field. But it is not clear how to introduce gravity when the particle is virtual (the cell rule of the cellular automaton should implement it).

\end{itemize}


\paragraph{Acknowledgments.}
Many thanks to Guillaume Boussinot for his remarks, criticisms, and suggestions.


\bibliographystyle{plain}
\bibliography{paper}

\end{document}